\documentclass[twocolumn,showpacs,preprintnumbers,amsmath,amssymb]{revtex4}
\usepackage{graphicx,epsf}
\begin{document}
\title{Jamming and asymptotic behavior in competitive random parking of bidisperse cars}
\author{M. Kamrul Hassan$^{\dagger,\ddagger}$, J\"urgen Schmidt$^\dagger$, Bernd Blasius$^\dagger$, and 
J\"urgen Kurths$^\dagger$}
\affiliation{
$^\dagger$University of Potsdam, Department of Physics, Postfach 601553, D-14415 Potsdam, Germany\\
$^\ddagger$ University of Dhaka, Department of Physics, Theoretical Physics Division, Dhaka 1000, Bangladesh
}
\date{\today}


\begin{abstract}%

We propose a generalized car parking problem where either a car of size $\sigma$ or of size $m\sigma$ ($m>1$) is 
sequentially parked on a line with probability $q$ and $(1-q)$, respectively. The free parameter $q$ interpolates 
between the classical car parking problem at either extreme ($q=0$ and $q=1$) and the competitive random sequential 
adsorption of a binary mixture in between.  We find that the coverage in the jamming limit for a mixture 
always exceeds the value obtained for the uni-sized case. The introduction of a bidisperse mixture results in the
slow approach ( $\sim t^{-1}$) to the jamming limit by the smaller species while the larger species reach 
their asymptotic values exponentially fast $\sim t^{-1}e^{-(m-1)qt}$.
   
\end{abstract}


\maketitle

The adsorption of particles on a flat substrate is a common phenomenon and it is of wide interest in physics, chemistry, 
biology  and in many other branches
of science and technology. Examples are the adsorption of macromolecules, colloidal particles, bacteria, proteins and 
latex particles \cite{kn.has1,kn.has2,kn.has3}. Due to its wide range of applications, there has been continuing effort
to develop and enrich our understanding of the topic through all the avenues of research comprising 
experimental, numerical and analytical means \cite{kn.has11}. 

From the theoretical viewpoint, the simplest model that can still capture the generic features of the 
adsorption phenomenon is the kinetics of {\it random sequential adsorption} (RSA) of monodisperse particles on a clean
 d-dimensional substrate \cite{kn.has4}. In this process, particles are  deposited randomly, one at each time step, with
 the restriction that overlap is forbidden. The one dimensional continuum version of the RSA process is 
popularly known as the {\it random car parking} (RCP) problem. This simple RSA model ignores diffusional relaxation, 
 precise particle-particle and particle-substrate interaction (though the hard-core interaction is incorporated)
and desorption or evaporation of the adsorbed particles. Nevertheless, there has been a continued 
effort to bridge the gap between the experimental conditions and the simplest form of RSA. As a result,
 there are many interesting variants of RSA. Examples include the cooperative sequential adsorption 
\cite{kn.has4,kn.has15}, the accelerated random sequential adsorption 
 \cite{kn.rodgers}, the ballistic deposition  \cite{kn.talbot}, the RSA on disordered substrates 
\cite{kn.naim}, the RSA of growing objects \cite{kn.rodgers1}. Several attempts have been made to include the 
transport of deposited particles by diffusion (see \cite{kn.privman} and references therein). 
One issue inherent to the experimental realization is that the adsorbing particles are not always perfectly monodisperse.
 Polydispersity is almost an inevitable property in many real life situations. The binary mixture is the simplest 
and the first step toward the understanding of polydisperse system. There has been an 
increasing interest in the study of the RSA of mixtures with a strict lower and upper size cut-off 
 \cite{kn.meakin,kn.bonnier,kn.has5,kn.has6}, and continuous mixtures obeying a power-law size distribution \cite{kn.has7}. It is important to note that in the former case the 
structure of the monolayer in the long time limit is described by the jamming coverage, whereas in the latter case 
the resulting monolayer is uniquely characterized by the fractal dimension of the arising pattern \cite{kn.has7}. Much of the theoretical effort was restricted only to a binary mixture of particles 
with very large size differences \cite{kn.has5,kn.has6}. However, mixtures with a  
narrow distribution of sizes have more practical relevance than mixtures with large size differences. 

In this article we consider the RSA of two species and incorporate a parameter that controls the competition between the 
adsorbing species. Hence we shall call it {\it competitive random sequential adsorption} (CRSA) \cite{kn.hassan}. 
The motivation of setting the present 
problem came from its potential applications in polymer, aerosol and colloidal science. Consider the case of the 
adsorption of polymer chains onto a surface or the reactions along polymer chains, which was in fact the motivation of 
the original work by Flory \cite{kn.has2}. The polymer chains are naturally not all of the same length, instead they 
are a mixture of chains of lengths differing by an integer multiple of a certain size unit. Of course, the ideal case 
to consider would be the study of a polydisperse mixture with strictly lower and upper size cut-off which would warrant 
greater complexity. Nevertheless, the present work still captures all the essential and generic aspects of the 
adsorption of mixtures in the simplest possible way. Thus, our model provides useful insight for the analysis of 
more complex problems. We find that the mixtures exhibit a behavior which is qualitatively and quantitatively 
different from that of the monodisperse adsorption.

It is worth to mention that although the RSA problem is based on a simplified picture, yet it is 
analytically tractable only in one dimension. This is typical 
for many problems of statistical physics. The underlying reason is that the 
{\it shielding property} can only be found in 1d. Here shielding means that in 1d the adsorption of a particle 
on an empty interval separates it into two disconnected smaller intervals, both having the same geometry as their 
parent interval and hence they can be treated as an independent interval. Therefore, much of the information in 
higher dimensions is provided by numerical simulations or by experiment. We can identify two separate 
characteristic time scales in adsorption phenomena: the time between depositions, $t_d$, and the time required 
for the deposited particles to reorganize themselves on the substrate, 
$t_r$. In the context of the present work, we restrict ourselves to the case where the relaxation time is very large, 
$t_d<<t_r$, so that the deposited particles on the substrate do not have enough time to reorganize to find the 
minimum energy configuration and hence the system is driven out of equilibrium. Such a situation 
is expected when the attractive contact energy between the adsorbed particles and the substrate is sufficiently higher 
than the thermal agitation energy $\sim k_BT$. Restriction to the monolayer formation and the 
irreversibility of the process 
make it one of the simplest yet interesting and challenging problems in statistical physics. Despite the inherent
simplicity of the RSA model, it has in fact reproduced many experimental results.
Namely, the RSA of monodisperse hard particles was first shown to describe the experimental results of the adsorption of
proteins \cite{kn.feder}, and later that of latex and submicron colloid particles \cite{kn.onada} on flat surfaces,
for example. In these processes, it is known that the relaxation is typically very slow and hence the simple RSA
algorithm works well. The known key features of the RSA processes are: 
\begin{itemize}
\item[{\it a)}] The process results in a unique jamming coverage in the thermodynamic limit. The final monolayer is 
uniquely characterized by the jamming 
coverage $\theta_d(\infty)$ in d dimensions, defined as the fraction of the substrate covered by the depositing 
particles, which 
is always less than unity due to its irreversible character, and the restriction to the monolayer growth. 
\item[{\it b)}] Near the jamming limit, the approach of the coverage $\theta_d(t)$ to its final value 
$\theta_d(\infty)$ obeys the relation 
\begin{equation}
\label{eq:feder}
\theta_d(\infty)-\theta_d(t) \sim t^{-1/d}.
\end{equation}
This is widely known as Feder's law \cite{kn.feder} and the theoretical arguments  supporting it have been presented 
independently by Swendsen and Pomeau \cite{kn.swendsen}. 
\item[{\it c)}] Another surprising result in RSA of d-dimensional hyperspheres is that an estimate of the final jamming 
coverage $\theta_d(\infty)$ can be obtained with a reasonably good degree of accuracy from the result of the one 
dimensional model (RCP) through the relation
\begin{equation}
\label{eq:palasti}
\theta_d(\infty)\approx (\theta_1(\infty))^d.
\end{equation}
Although the strict equality as conjectured by Palasti \cite{kn.palasti} is not valid, the relation often provides a remarkably 
accurate estimate. 
\end{itemize}

The problem of the competitive random sequential adsorption can be formulated as follows. Assume that we have a particle reservoir which contains a binary mixture of chemically identical 
particles of two different sizes. Further, assume that a fraction $q$ of the particles in the reservoir is of size 
$\sigma$ and the rest, i.e.\ a fraction $(1-q)$ of size $m\sigma$, $m>1$. The algorithm of the process is then as follows
\begin{itemize}
\item[{\it i)}] A particle is picked randomly for deposition on the substrate: it can be of size $\sigma$ with 
probability $q$ or of size $m \sigma$ with probability $p=1-q$.
\item[{\it ii)}] Randomly a position is chosen on the substrate.
\item[{\it iii)}] The particle is deposited on the substrate if the chosen position falls on an empty interval
and it is large enough to accommodate the particle. Otherwise the particle is rejected.
\item[{\it iv)}] In either case the time is increased by one unit. 
\item[{\it v)}] The process is repeated until no more gaps with size larger than $\sigma$ are left on the substrate.
\end{itemize}

The task now rests on translating the problem in question into a set of equations for the gap size distribution function 
$G(x,t)$ to describe how it evolves in time. Before doing so, we first note that the two events (adsorption of $\sigma$ 
and $m\sigma$) are mutually exclusive, i.e.\ at each time step only one particle is adsorbed. This is ensured by the 
fact that either the particle of size $\sigma$ is adsorbed with probability $q$ or a particle of size $m\sigma$ is 
adsorbed with probability $(1-q)$. Then, we can immediately generalize the classical car parking problem and write the 
following set of  equations for $G(x,t)$ for $x\geq m\sigma$ 
\begin{eqnarray}  
\label{eq:xgtmsig}
{{\partial G(x,t)}\over{\partial t}} & = & -(x-\sigma_{av})G(x,t)+ 2q\int_{x+\sigma}^\infty G(y,t)dy \nonumber \\  &+ &
2p\int_{x+m\sigma}^\infty G(y,t)dy,
\end{eqnarray}
and for $x<m\sigma$  
\begin{eqnarray}
\label{eq:xltmsig}
{{\partial G(x,t)}\over{\partial t}} & = & 2q \int_{x+\sigma}^\infty G(y,t)dy + 2p\int_{x+m\sigma}^\infty G(y,t)dy
 \nonumber \\ &-& H(x-\sigma)q(x-\sigma)G(x,t),
\end{eqnarray}
where $\sigma_{av}=\{m+q(1-m)\}\sigma$ is the average particle size and $H(z)$ is the Heaviside step function. 
Equations (\ref{eq:xgtmsig}) and (\ref{eq:xltmsig}) 
satisfy the conservation law 
\begin{equation}
\label{eq:cons}
\int_0^\infty \Big(x+\{m+q(1-m)\}\sigma \Big)G(x,t)dx=1.
\end{equation}   
To solve the kinetic equation, we seek in Eq.\ (\ref{eq:xgtmsig}) a trial solution of the form 
\begin{equation}
\label{eq:g}
G(x,t) = A(t) e^{-(x-\sigma_{av})B(t)}\,,
\end{equation}
where $A(t)$ and $B(t)$ are yet undetermined functions. Substituting the trial solution into Eq.\ (\ref{eq:xgtmsig}), 
we obtain the following two differential equations for $A(t)$ and $B(t)$ 
\begin{equation}
{{d \ln A(t)}\over{dt}}=2q{{e^{-\sigma t}}\over{t}}+2p{{e^{-m\sigma t}}\over{t}}, \hspace{4mm} {{dB(t)}\over{dt}}=1.
\end{equation}
Solving these equations subject to the initial conditions
\begin{equation}
 \lim_{L\longrightarrow \infty}\int_0^\infty G(x,0)dx =0, 
\lim_{t \longrightarrow 0}\int_0^\infty xG(x,t)dx=1,
\end{equation}
and satisfying the conservation law Eq.\ (\ref{eq:cons}), gives  
\begin{equation}
\label{eq:a}
A(t)=t^2F(\sigma t) \hspace{4mm} {\rm and} \hspace{4mm} B(t)=t,
\end{equation}
where the dimensionless quantity $F(\sigma t)$ is
\begin{equation}
F(\sigma t)= e^{-2\int_0^{\sigma t}\{q(1-e^{-u})+p(1-e^{-mu})\}du/u}.
\end{equation} 
Note that if we set $\sigma=0$ in Eq.\ (\ref{eq:xgtmsig}), we can still recover the well known solution of random 
scission model of a binary fragmentation process $G(x,t)=t^2\exp[-xt]$ \cite{kn.has9}.
The complete solution to Eq.\ (\ref{eq:xgtmsig}) is therefore
\begin{equation}
\label{eq:soll}
G(x,t)=t^2 F (\sigma t)e^{-(x-\sigma_{av})t}  
\hspace{3mm} {\rm for}  \hspace{3mm}  x \geq m\sigma.
\end{equation}
We can solve Eq.\ (\ref{eq:xltmsig}) directly for a narrow size difference ($1<m\leq 2$), since in this case only the known solution (\ref{eq:soll}) contributes to the integrals in (\ref{eq:xltmsig}). To this end we seek a trial solution of the form
\begin{equation}
G(x,t)=C(x,t)e^{-q(x-\sigma)t}.
\end{equation} 
In this domain, $\sigma$ is the minimum gap size, and hence the minimum value of the lower limit of the first integral 
of Eq.\ (\ref{eq:xltmsig}) is $2\sigma$. Therefore, if  we restrict to $1<m\leq 2$ 
we can substitute the solution of 
Eq.\ (\ref{eq:xgtmsig}) into both integrals and obtain the solution for $\sigma<x<m\sigma$
\begin{eqnarray}
G(x,t) &=& \int_0^t F(\sigma s)s(2qe^{\sigma(mp-1)s}+2p e^{-qm\sigma s})\nonumber \\ & \times &
e^{-(x(qt+ps)-q\sigma t)}ds.
\end{eqnarray}
Defining $G(x,t)\equiv G(x-\sigma_{av},t)$ for $ x \geq m\sigma $ we find that the solution of Eq.\ (\ref{eq:xgtmsig}) 
satisfies the identity 
\begin{equation}
G(\{x-\sigma_{av}\}\lambda, t/\lambda)=\lambda^{-2}G(x-\sigma_{av},t).
\end{equation}
This relation is the hallmark for the existence of scale invariance and it is equivalent to the {\it data-collapse} 
formalism \cite{kn.stanley}. It means that if the deposition rate $(x-\sigma_{av})$ is increased by a factor $\lambda$, and the 
observation time is decreased by the same factor, then the resulting structure would look the same except for the 
numerical pre-factor.

From the gap size distribution function one can immediately find the total coverage $\theta_T(t)$ by the 
adsorbed particles through the definition 
\begin{equation}
\label{eq:jamtot}
\theta_T(t)=1-\int_0^\infty x\,G(x,t)dx.
\end{equation}  
For $0<q<1$, both species contribute to the total coverage $\theta_T(t)=\theta_S(t)+\theta_L(t)$. In order to obtain expressions for the coverages by the small and the large particles, $\theta_S(t)$ and $\theta_L(t)$, we 
substitute the solutions of Eq.\ (\ref{eq:xgtmsig}) and (\ref{eq:xltmsig}) in $\dot\theta_T(t)$. After some lengthy yet simple manipulations, we obtain the 
contribution of the small particles $\theta_S(t)$
\begin{eqnarray}
\label{eq:jam_small}
\theta_S(t)  &=&   2 \int_0^t ds F(s)se^{kps}(qe^{-s}+pe^{-(k+1)s})\nonumber \\ &&
\times \Big({{e^{-k(ps+qt)}-1}\over{(ps+qt)}} -{{e^{-ks}-1}\over{s}}\Big)\nonumber \\ && +q \int_0^t F(s)(1+ks)e^{-kqs}ds,
\end{eqnarray}
where $k=(m-1)$,
 and the contribution of the large particles $\theta_L(t)$
\begin{equation}
\label{eq:jam_large}
\theta_L(t)=pm\int_0^tF(s)e^{-kqs}ds\,.
\end{equation}
\begin{figure}[!htb]
\centerline{\includegraphics[width=8.0cm]%
{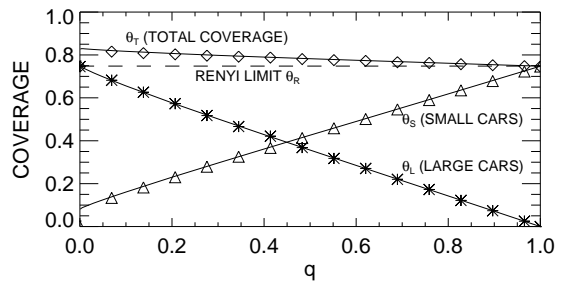}}
\caption[]{Coverage in the jamming limit vs $q$ for the respective contributions from small and large 
particles for the case $m=1.5$. Symbols represent simulational results, lines the analytical solution (\ref{eq:jam_small},\ref{eq:jam_large}).
 Simulations were carried out for a substrate length of 10000 small cars, each symbol is an average over 20 independent 
realizations.} 
\label{fig:1}
\end{figure}
We calculate the coverage in the jamming limit ($t\longrightarrow \infty$) of our 
model by direct numerical integration of the 
above two equations. The two terms in $\theta_S$ are the contributions made by the adsorption of small particles in 
the domains $\sigma\leq x<m\sigma$ and $x\geq m\sigma$, respectively. 
The individual concentration of cars of
size $\sigma$ and $m\sigma$ on the monolayer depends on their respective probabilities $q$ and $p$. The final 
concentration of small particles will always exceed $q$, since the smaller particles have a higher success rate
in the competition with the larger species, in particular when the substrate is already densely populated. 
For $m=1$, the first term in $\theta_S(t)$ 
is zero and the remaining terms of $\theta_T(t)$ yield
\begin{equation}
\theta_T(t)=\int_0^t F(s)ds=\theta_{{\rm R}}(t),
\end{equation}
where $\theta_{{\rm R}}(t)$ is the coverage for RCP model with $\theta_{{\rm R}}(\infty)=0.748...$, the 
famous Renyi-limit. 

To test our analytic results, we simulate the CRSA problem on a computer. Naturally, the simulations restrict on  
finite substrate lengths $L$. However, for sufficiently large lengths $L\gg\sigma$ the effects of finite $L$ are small. 
The algorithm we use follows literally the steps {\it i)} -- {\it v)} described in the introduction. The time scale in 
the simulations is set by the number of particles brought to the substrate (if adsorbed or not). 
The discrete `loop index' $i$ of the simulation relates to the time scale of the rate equations $t$ as
\begin{equation}
t = i \frac{\sigma}{L}\,.
\end{equation}

 \begin{figure}[!htb]
\centerline{\includegraphics[width=8.0cm]%
{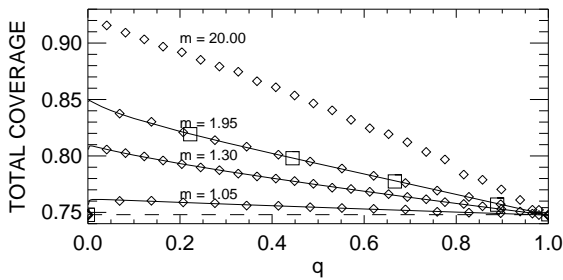}}
\caption[]{The total coverage vs $q$ for various values of $m$. Diamond symbols refer to simulations with a substrate 
length of 10000 small cars, and for the case $m = 1.95$ results from a simulation with $L = $ 5000 are shown for 
comparison as square symbols..} 
\label{fig:2}
\end{figure}

Using this time scale and expressing all lengths in the simulation in terms of $\sigma$, the simulational results can 
be directly compared to the solution to the rate equations. In Fig.\ \ref{fig:1} the 
results for the jamming coverage are shown for the case $m = 1.5$, where the parameter $q$ is varied. In general, we 
find an excellent agreement between simulation and our theory. The total coverage is found to be always larger for binary 
mixtures ($q \ne 0,\,q\ne 1$) than for the uni-sized case ($q = 0$ or $q = 1$).  
For a given $m$ the maximum coverage $\theta_T^{{\rm max}}(\infty)$ occurs when the concentration of the smaller species in the 
reservoir is small, i.e.\ at $q\longrightarrow 0^+$. In this limit, the larger species are adsorbed first 
until their jamming limit is reached. Then the remaining gaps are filled by the smaller species. In other words, 
the two processes (adsorption of small and large particles) are decoupled. As a result, the chronology of the adsorption 
of the small and large species strongly influences the dynamics of the system {\it vis-a-vis} the final coverage as
well as the final number density of each species on the substrate. 
Note that the particles are drawn from an infinite reservoir, i.e.\ no 
matter how small the concentration of small particles in the reservoir will be, they still may be picked up at the 
expense of many failures. The maximum coverage for a given $m$ occurs at $q\longrightarrow 0^+$ and it decreases
monotonically as $m$ decreases, eventually coinciding with the Renyi-limit at $m=1$. The highest 
coverage is obtained for a large size ratio ($m\longrightarrow \infty$) 
of the binary mixture (see Fig.\ \ref{fig:2}).  In the limit $m\longrightarrow \infty$ and $\sigma>0$
the maximum total coverage is 
\begin{equation}
 \theta_T^{{\rm max}}(\infty)= \theta_{{\rm R}}(\infty)+(1-\theta_{{\rm R}}(\infty))\theta_{{\rm R}}(\infty)=
0.937\ldots\,.
\end{equation} 
The simulation with $m = 20$ shown in Fig.\ \ref{fig:2} is already very close to this limit. This implies that a
maximum of $93.7$ percent of the total space can be covered by a randomly adsorbed binary mixture. Note that the coverage
 for the CRSA of a binary mixture has no smooth transition to that of a mixture of points and 
finite sized particles in the limit $m\longrightarrow \infty$; the small particles, no matter how small their finite 
length is, will always 
contribute to $\theta_S$, and thus, the total 
coverage $\theta_T$ is always larger than the Renyi-limit $\theta_{{\rm R}}(\infty)$. Only if their length is indeed zero, so 
that $\theta_S(\infty)=0$, the only role the point particles can play is to prevent the large particles from reaching the 
Renyi-limit. Also this problem may be solved exactly \cite{kn.has10}. The result for the total coverage
\begin{equation}
\theta_T(t)=p\int_0^tF(s)e^{-qs}ds.
\end{equation}
In this case the total coverage stems from the large particles alone and Feder's law does no longer holds.
\begin{figure}[!htb]
\centerline{\includegraphics[width=8.0cm]%
{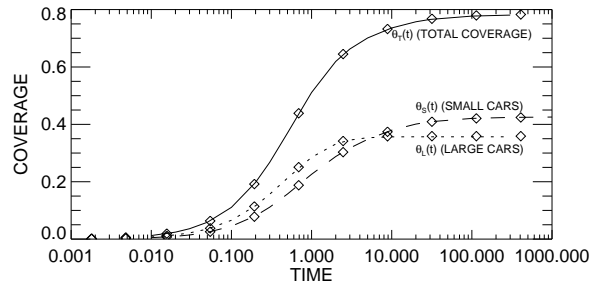}}
\caption[]{Time development of the coverage for $q=0.5$, $m = 1.5$. Lines (solid, dashed, dotted) are theoretical results,
 symbols from simulations.} 
\label{fig:3}
\end{figure}

From the gap size distribution function one can also obtain the number density of the individual species, using the 
definition 
\begin{equation}
n_S+n_L=\int_0^\infty dyG(y,t). 
\end{equation}
Using similar manipulations as we did for the jamming coverage we find 
\begin{equation}
n_S(t)=\theta_S(t) \hspace{1.cm} {\rm and} \hspace{1.cm} n_L(t)=\theta_L(t)/m.
\label{eq:numberdensity}
\end{equation}

\begin{figure}[!htb]
\centerline{\includegraphics[width=8.0cm]%
{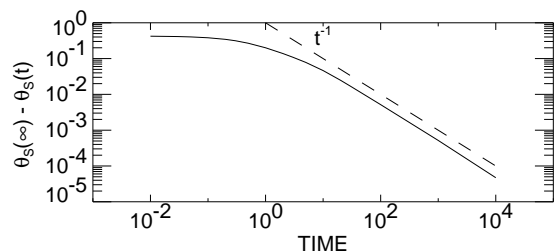}}
\caption[]{The approach of $\theta_S(t)$, and thus $\theta_T(t)$, to its asymptotic value obeys Feder's law (equation 
(\ref{eq:feder})).} 
\label{fig:4}
\end{figure}

Fig.\ \ref{fig:3} shows the time dependence of the coverage by small, large, and by 
both species. $\theta_L(t)$ approaches its asymptotic value (the jamming limit) much more rapidly 
than $\theta_S(t)$. In order to quantify the approach to the respective asymptotic values we plot 
$\ln[\theta_S(\infty)-\theta_S(t)]$ vs $\ln[t]$ (Fig.\ \ref{fig:4}). The 
small particles reach their 
jamming limit algebraically, obeying Feder's law (Eq.\ (\ref{eq:feder})). Asymptotically the coverage for the 
large particles, Eq.\ (\ref{eq:jam_large}) behaves as
\begin{equation}
\label{eq:largeasym}
\theta_L(t) \sim \theta_L(\infty)-e^{-(m-1)qt}/t.
\end{equation}
Thus the large particles reach their 
asymptotic value exponentially with an algebraic prefactor $t^{-1}$, 
hence not obeying Feder's law. This implies that the asymptotic behavior of the total coverage is dominated by 
the slow dynamics of the smaller specie, approaching the 
final values algebraically as $t^{-1}$ (see Fig.\ \ref{fig:4}). Thus, we verified the validity of Feder's law for the 
present system. Note that according to Eq. (\ref{eq:numberdensity}) the number densities will follow the
asymptotic behavior of the respective coverages.  

The present work provides an exact theoretical basis for the RSA of mixtures. Our results are in good qualitative 
agreement with the numerical simulations presented by Meakin and Jullien \cite{kn.meakin} for the {\it two-dimensional} 
CRSA of disks. Especially, they find a more efficient coverage for the mixture as well. This is in contrast to the study 
of CRSA by Bonnier \cite{kn.bonnier} in 1d, who claims that this behavior was typical for lattice models, while his 
continuum model predicts a less efficient coverage by the mixture than by the RSA of uni-sized objects.
The continuum model we have studied in this work is consistent with its lattice counterpart. Our results are supported by 
the good agreement with the direct simulation. Furthermore, the dependence of our results on $q$ and $m$ is in excellent qualitative agreement with the results reported in \cite{kn.meakin}. Further, Meakin and Jullien too confirm 
Feder's law for the small particle species, and find an exponential approach of the coverage by the large ones. 
In this context it is interesting to verify Palasti's conjecture (Eq.\ (\ref{eq:palasti})), taking the values from 
Fig.\ 3 of the paper by Meakin and Jullien for the jamming coverage by a binary mixture of disks. We find a deviation 
of at most 5\% from the estimates provided by our one-dimensional data and Eq.\ (\ref{eq:palasti}). 
This confirms that 1d models, despite their simplicity, indeed may prove useful in gaining information about higher dimensions, in contrast to many problems in equilibrium statistical physics. 
There, in general, 1d systems often have very little relevance to 2d or to higher dimension, e.g.\ no 
thermodynamic phase transitions are possible in 1d.
In RSA 
there is no phase transition in the adsorbed 
monolayer regardless of its dimensionality. The 1d model already has all the ingredients such as total irreversibility 
with the associated memory effect or exclusion or blocking phenomena that make their behavior nontrivial and 
qualitatively similar to that observed in higher dimensions, except a trivial quantitative difference. Of course, 
the RSA with hard objects covers less and less fraction of the substrate as the dimension increases. 

In conclusion, we have shown that an analytical model for the CRSA of a 
binary mixture in one dimension predicts a jamming limit that is always larger than that for the uni-sized case. In the 
limits $q=0$, $q=1$, as well as $m=1$ we recover the classical result $\theta_{{\rm R}}=0.748$ for the equal 
sized RSA process. 
In all other cases, i.e.\ for true binary mixtures, we find a larger jamming limit in theory and simulation. We have 
explicitly shown that not only the size ratio between the two species $m$ which is an important parameter, but also 
the concentration of the individual species in the reservoir which depends on $q$ is an equally important parameter. 
It appears that the smaller the 
concentration of the small particles in the reservoir, the higher the jamming coverage. 
In general, our analytical results are in excellent agreement with direct numerical simulations. In addition, they show
an excellent qualitative agreement with numerical simulation results in higher dimension which emphasize the 
importance of 1d model. The time-asymptotic approach of the jamming limit is dominated by the contribution of the 
small particles, and we confirm the $1/t$ behavior predicted by Feder's law in one dimension. The large particles reach their contribution 
to the jamming limit exponentially with an algebraic pre-factor.

This work was supported by the  Alexander von Humboldt foundation under the Georg Forster Fellowship (M.\ K.\ H.), 
by the Ministerium f\"ur Forschung in Brandenburg (J.\ S.), and by the Deutsche Volkswagen-Stiftung (B.\ B.).

\end{document}